\documentstyle[12pt]{article}
\newcommand{\mn}{^{\mu\nu}}
\newcommand{\Tf}{T_f}
\newcommand{\Tq}{T_q}
\newcommand{\Th}{T_h}
\newcommand{\Tc}{T_c}
\newcommand{\Tr}{T_{\rm reheat}}
\newcommand{\vs}{v_{\rm shock}}
\newcommand{\vf}{v_{\rm fluid}}
\newcommand{\vd}{v_{\rm def}}
\newcommand{\vh}{v_h}
\newcommand{\vq}{v_q}
\newcommand{\pq}{p_q}
\newcommand{\ph}{p_h}
\newcommand{\wq}{w_q}
\newcommand{\wh}{w_h}
\newcommand{\aq}{a_q}
\newcommand{\ah}{a_h}
\newcommand{\sh}{s_h}
\newcommand{\sq}{s_q}
\newcommand{\la}[1]{\label{#1}}
\newcommand{\beq}{\begin{equation}}
\newcommand{\eeq}{\end{equation}}
\newcommand{\bea}{\begin{eqnarray}}
\newcommand{\ba}{\begin{eqnarray*}}
\newcommand{\eea}{\end{eqnarray}}
\newcommand{\ea}{\end{eqnarray*}}
\newcommand{\tyhja}{\vspace{0.5cm}}

\newcommand{\eg}{e.g.\ }

\newcommand{\eqs}{eqs.~}
\newcommand{\fig}{fig.~}

\newcommand{\oder}[2]{{\frac{\partial #1}{\partial #2}}}

\let\oldref=\ref
\renewcommand{\ref}[1]{(\oldref{#1})}
\newcommand{\doo}{\partial}
\newcommand{\bfx}{{\bf x}}
\newcommand{\bfv}{{\bf v}}
\newcommand{\vdef} {v_{\rm def}}
\newcommand{\vdet} {v_{\rm det}}
\newcommand{\ew} {{\bf ew}}
\newcommand{\qh} {{\bf qh}}

\begin{document}
\setcounter{page}{0}
\setlength{\parindent}{0.0cm}
\title{ }
\author{\hspace*{-0.8cm}
{\bf J. Ignatius\thanks{\twlrm email: ignatius@phcu.helsinki.fi}$^{\;\, ,b}$,
K. Kajantie\thanks{\twlrm email: kajantie@phcu.helsinki.fi}$^{\;\, ,a}$,
H. Kurki-Suonio\thanks{\twlrm email: hkurkisu@pcu.helsinki.fi}$^{\;\, ,b}$
and M. Laine}\thanks{\twlrm email: mlaine@phcu.helsinki.fi}$^{\;\, ,a}$}
\date{ }
\maketitle
\begin{center}
\vspace*{-4cm}
{\Large\bf THE GROWTH OF BUBBLES IN} \\
\vspace*{2mm}
{\Large\bf COSMOLOGICAL PHASE TRANSITIONS} \\
\vspace*{13mm}
{\sl  $\mbox{}^{a}$Department of Theoretical Physics and \\
$\mbox{}^{b}$Research Institute for Theoretical Physics, }
\\
{\sl  P.O. Box 9, 00014 University of Helsinki, Finland}
\\
\vspace*{0.3cm}
27 September 1993
\end{center}
\vspace*{-9cm}
\hfill Preprint HU-TFT-93-43
\vspace*{9cm}
\begin{center}
{\large\bf Abstract}\\
\vspace*{5mm}
\parbox{14.5cm}{
We study how bubbles grow after the initial nucleation event in
generic first-order cosmological phase transitions characterised
by the values of latent heat~$L$, interface tension~$\sigma$ and
correlation length~$\xi$, and driven by a scalar order
parameter~$\phi$. Equations coupling~$\phi(t,\bfx)$ and the fluid
variables~$\bfv(t,\bfx)$, $T(t,\bfx)$ and depending on a dissipative
constant~$\Gamma$ are derived and solved numerically in the 1+1 dimensional
case starting from a slightly deformed critical bubble
configuration~$\phi(0,\bfx)$. Parameters~$L$, $\sigma$, $\xi$
corresponding to QCD~and electroweak phase transitions are chosen and
the whole history of the bubble with formation of combustion and shock
fronts is computed as a function of~$\Gamma$. Both deflagrations and
detonations can appear depending on the values of the parameters.
Reheating due to collisions of bubbles is also computed.\\
\vspace*{1mm}

}
\end{center}
\newpage
\setcounter{page}{1}
\setlength{\parindent}{1.2cm}

\section{Introduction}
\la{sec:intro}

The bubble nucleation in cosmological first-order phase transitions
as well as the propagation and stability of planar interfaces have been
discussed extensively in the literature~\cite{coleman}--\cite{adams}.
The purpose of this paper is to join these two ends of the life of a bubble
by giving a dynamical model which describes the entire
history of the bubble from the initial configuration
via initial acceleration into a large bubble growing with
constant velocity.

The stage for the events in this paper is the cosmic fluid
with a first-order phase transition at $T=T_c$ --- in
practice either the QCD or the electroweak (EW) phase
transition. For physical quark masses there is no
symmetry associated with the former one and its order is
not definitely confirmed \cite{karsch}. The latter is
a symmetry breaking transition and, at least for not too
large Higgs masses, it is of first order both on the basis
of perturbative \cite{dj}--\cite{hebecker} and nonperturbative
lattice Monte Carlo \mbox{\cite{bunk}--\cite{fkrs}} work.
Inflationary transitions are essentially vacuum ones and
the considerations here do not apply.

The main quantity characterizing the cosmic fluid is
its energy-momentum tensor $T^{\mu\nu}=(\epsilon+p)
u^\mu u^\nu-p g^{\mu\nu}$. The first-order nature
implies that there are two phases, a high temperature
phase (symmetric, quark-gluon plasma phase)
with pressure $p_q(T)$ and a low temperature phase
(broken symmetry, hadron phase) with pressure $p_h(T)$, which
can coexist at $T_c$: $p_q(T_c)=p_h(T_c)$ but
$p_q'(T_c)>p_h'(T_c)$. We shall describe the transition by
a scalar order parameter field $\phi(t,\bfx)$. This is
obvious for a symmetry breaking transition, but we shall
use the same description also if no symmetry is involved:
the order parameter could then be, for example, the energy
or entropy density. The bubbles
are configurations of $\phi(t,\bfx)$.

The problem now is to derive equations of motion for
the total cosmic fluid~-- order parameter field system.
We carry this out in analogy
with the reheating problem in inflation \cite{reheating,hannu1}.
The total energy momentum tensor of the system is
conserved ($\partial_\nu T^{\mu\nu}=0$) but those of
the fluid and order parameter field subsystems are not.
Physically, entropy produced at the bubble interface
couples the behavior of~$\phi$ with the fluid. The
strength of this coupling is described by
a dissipative constant~$\Gamma$. It has
been related by a fluctuation-dissipation formula
to equilibrium averages in refs.~\cite{khlebnikov,arnold},
but we use $\Gamma$ as a phenomenological parameter.
Estimates for it in the~EW theory have been given in
refs.~\cite{turok,ck}.

Given the equations of motion one can solve them numerically
and study how bubbles corresponding to given initial
supercooling (which follows from nucleation analysis),
given parameters of the transition (latent heat, interface
tension, correlation lengths), and given dissipative
constant $\Gamma$ evolve. Collisions of bubbles can be
similarly studied. In this first discussion we shall solve
the equations in 1+1 (one time, one space) dimensions,
which contains the main qualitative features. For complete quantitative
results one must go to 1+2 or 1+3 dimensions.

The initial stages of bubble growth have also been computed
by integrating
spherically symmetric
1+3 dimensional hydrodynamic equations in ref.~\cite{pantano}.
The velocity is taken as a free parameter,
effectively parametrised by the magnitude of energy flux.
Our work differs in one essential aspect from this:
the dissipative constant $\Gamma$ is at least in principle
calculable from the theory.


The paper is organized as follows. In Section~\oldref{sec:model}
we introduce our model
for the first-order phase transitions. In Section~\oldref{sec:defldeto}
we review the general hydrodynamic conditions for the bubbles, and
describe the two kinds of solutions, detonations and deflagrations.
In Section~\oldref{sec:time-dep} we describe our
results for different time-dependent phenomena like the initial
stages of bubble formation, the sharpening of shock fronts and
the collisions between expanding bubbles. In Section~\oldref{sec:stationary}
we give an account of the different steady-state variables for
deflagrations (temperatures and velocities) as
a function of $\Gamma$. The conclusions are in Section~\oldref{sec:concl}.

\section{Equations of motion for the cosmic fluid and the order
parameter field}
\label{sec:model}

The system we consider contains the cosmic fluid which has
supercooled in the metastable high $T$ phase (call it by
convention q) to some temperature $\Tf$. At this temperature
nucleation of bubbles of the low $T$ phase (call it by convention
h) becomes sufficiently frequent for the phase transition to
effectively take place. We shall first define the quantities
appearing in the equations of motion.

The bubbles are defined as configurations $\phi(t,\bfx)$ of a
scalar order parameter~$\phi$. The (meta)stable states of
the system are defined by the minima of the
effective potential $V(\phi ,T)$ of the order parameter $\phi$.
The equations will be formulated for a general~$V$,
but for numerical calculations we shall use a quartic
parametrisation~\cite{linde1,linde2}:
\beq
V(\phi,T)={1\over2}\gamma(T^2-T_0^2)\phi^2
-{1\over3}\alpha T\phi^3+{1\over4}\lambda\phi^4.
\la{V}
\eeq
The full functional form of the ring summation improved
effective potential~\cite{carrington}--\cite{hebecker}
deviates somewhat from this, and even more does the effective potential
containing nonperturbative effects~\cite{fkrs}. The methods
developed here can be straightforwardly extended to these
improved potentials.

The physical quantities corresponding to the parameters
$\alpha$, $\gamma$, $\lambda$ and $T_0$ in eq.~\ref{V}
are the latent heat $L$, the interface tension $\sigma$, the critical
temperature $T_c$ and the correlation length
$\xi(T_c)\equiv \xi_c$ \cite{keijo}. The quartic parametrisation
in eq.~\ref{V} implies that the correlation lengths in the
two phases at $T_c$ are equal, but actually they are
different \cite{fkrs}.

The primary physical quantities characterizing the transition
are $T_c$, $L$, $\sigma$~and $\xi_c$, and given these, we can always
solve for the parameters
$\alpha$, $\gamma$, $\lambda$ and $T_0$ in eq.~\ref{V}. In
this way we can use eq.~\ref{V} also for QCD. Its use for
a symmetry breaking transition is, of course, obvious.

For the equation of state of the system we shall take
\beq
p_q(T)=aT^4 \qquad p_h(T)=aT^4+B(T), \la{eos}
\eeq
where $B(T)=-V(\phi_{\rm min},T)$ is the
difference between the free energy densities
of the symmetric and the broken symmetry phases.
The number of effective degrees
of freedom of the high temperature phase is denoted by $g_*$,
and $a\equiv (\pi ^2/90)g_*$.

Summarizing, our dynamical variables
are the four-velocity of the fluid $u^{\mu}(t,{\bf x})$,
the scalar field $\phi (t,{\bf x})$
and the local temperature $T(t,{\bf x})$. We use the notations
$w_r=4aT^4$ and $p_r=aT^4$
for the pure radiative enthalpy and pressure, respectively.

To motivate and to write down the equations of motion we
first write the effective potential of eq.~\ref{V}
in two parts:
\beq
V(\phi ,T)=\underbrace{-\frac{1}{2}\gamma T_0^2\phi ^2
+\frac{1}{4}\lambda\phi ^4}_{V_0(\phi)}+
\underbrace{\frac{1}{2}\gamma T^2\phi ^2-\frac{1}{3}
\alpha T\phi ^3}_{V_1(\phi ,T)}
\,\, . \la{twoparts}
\eeq
The part $V_0(\phi )$ is related strictly to the scalar field, whereas
the part $V_1(\phi ,T)$ includes the interaction of the field with the
thermal bath. The total energy-momentum tensor is
\bea
T\mn & = &\partial ^{\mu}\phi\partial ^{\nu}\phi
-g\mn [\frac{1}{2}\doo _{\alpha}\phi\doo ^{\alpha}\phi -V_0(\phi )]   \\
&   & \mbox{}+ [w_r-T\frac{\doo V_1(\phi ,T)}{\doo T} ]u^{\mu}u^{\nu}-
g\mn [p_r-V_1(\phi ,T) ]
\,\, , \la{Ttotal}
\eea
and it is of course conserved:
\beq
\doo _{\mu}T\mn =0
\,\, .\la{cons}
\eeq

The idea now is to bring in a dissipative term which acts as a friction
force for the scalar field and accounts for the entropy production
at the phase transition surface.
This is done by splitting eq.~\ref{cons} into two:
\beq
0=\doo _{\mu}T\mn=[\doo _{\mu}T\mn ]_{\phi}+
[\doo _{\mu}T\mn ]_{\rm rad}=\delta ^{\nu}-\delta ^{\nu}
\,\, .\la{split}
\eeq
The form of the Lorentz-covariant dissipative term $\delta ^{\nu}$ is adopted
from the context of inflation \cite{reheating,hannu1}.
Because of the temperature
dependence of the effective potential, the choice of
the terms $[\doo _{\mu}T\mn ]_{\phi}$
and $[\doo _{\mu}T\mn ]_{\rm rad}$ in eq.~\ref{split} is not
unique.
It seems most natural to make the splitting in the following way:
\beq
\left.
\begin{array}{l}
\doo _{\mu}\{\partial ^{\mu}\phi\partial ^{\nu}\phi
-g\mn [\frac{1}{2}\doo _{\alpha}\phi\doo ^{\alpha}\phi -V_0(\phi )]\}
+\oder{V_1(\phi ,T)}{\phi}\doo ^{\nu}\phi=-\frac{1}{\Gamma}
u^{\mu}\doo _{\mu}\phi\doo ^{\nu}\phi
\\
\doo _{\mu}\{[w_r-T\frac{\doo V_1(\phi ,T)}{\doo T} ]u^{\mu}u^{\nu}-
g\mn p_r\}+\oder{V_1(\phi ,T)}{T}\doo ^{\nu}T=+\frac{1}{\Gamma}
u^{\mu}\doo _{\mu}\phi\doo ^{\nu}\phi
\end{array}
\right. \la{yhta}
\eeq
After all, if $V_1(\phi ,T)$ could have been written in the form
$V_1(\phi ,T)=V_{\phi}(\phi)+V_T(T)$, there would be no
ambiguity in the splitting, and exactly \eqs\ref{yhta} would
result. Our final equations are then
\beq
\left.
\begin{array}{l}
\doo _{\mu}\doo ^{\mu}\phi+\oder{V}{\phi}=-\frac{1}{\Gamma}u^{\mu}
\doo _{\mu}\phi
\\
\doo _{\mu}\{[w_r-T\frac{\doo V}{\doo T} ]u^{\mu}u^{\nu}-
g\mn [ p_r -V]\}=(\frac{1}{\Gamma}
u^{\mu}\doo _{\mu}\phi+\oder{V}{\phi})\doo ^{\nu}\phi \,\, ,
\end{array}
\right. \la{yhtalot}
\eeq
where a common factor has been dropped.

Contracting both sides of the lower of \eqs\ref{yhtalot}
with the fluid four-velocity $u_{\nu}$ one gets
\beq
T\doo _{\mu}[(s_r-\oder{V}{T})u^{\mu}]=\frac{1}{\Gamma}
(u^{\mu}\doo _{\mu}\phi)^2 \,\, ,  \la{entropy}
\eeq
where $s_r=4aT^3$ is the radiative entropy and $-\oder{V}{T}$
the entropy associated with the order parameter.
This equation relates entropy production and the gradients of
$\phi$ via the constant $\Gamma$. Note that in a weak
coupling theory \cite{turok,ck} $1/\Gamma\sim 1/\tau_c \sim nv\sigma
\sim g^2T$.

In this paper we study \eqs\ref{yhtalot} in 1+1 dimensions,
which corresponds to planar symmetry. While this is a drastic
simplification, it nevertheless should correctly describe the late stages
of the bubble growth in the 1+3 dimensional world. Planar symmetry
also allows us to compare our results with analytical calculations.

In the planar-symmetry case it is also illuminating to
write down the equations for the steady-state solution. At
large times the system should evolve to a solution containing
a combustion front moving at constant velocity. In the rest
frame of the front all time derivatives then vanish and
eqs.~\ref{yhtalot} become
\beq
\phi ''(x)=\oder{V}{\phi}+\frac{v\gamma }{\Gamma}\phi '(x)
\la{stat1}
\eeq
\beq
(4aT^4-T\oder{V}{T})\gamma^2v={\rm const.}
\la{stat2}
\eeq
\beq
(4aT^4-T\oder{V}{T})\gamma^2v^2+aT^4+
\frac{1}{2}\phi '(x)^2-V={\rm const.}
\la{stat3}
\eeq
Eqs.~\ref{stat2} and \ref{stat3} actually also follow from
the steady-state energy-momentum conservation:
\beq
\doo _xT^{x\mu}=0
\,\, .
\eeq
Similarly, the entropy production equation \ref{entropy}
becomes
\beq
T{d\over dx}[(s_r-{\partial V\over \partial T})\gamma v]
={1\over\Gamma}\gamma^2v^2[\phi'(x)]^2.
\la{steadyentropy}
\eeq

The standard analysis of deflagration and detonation bubbles
\cite{gyulassy} is a study of what solutions eqs.~\ref{stat2},
\ref{stat3} allow. For given initial $T_f$ this leaves a one
parameter family of solutions.
For detonations there are four quantities, $T_q,v_q,T_h,v_h$,
constrained by two equations and by the boundary value $T_q=T_f$.
For deflagrations
the shock front has to be taken into account, and there are two more
quantities but also two more equations.
This is discussed in Section~\oldref{sec:defldeto}.
Eq.~\ref{stat1} is the new one which
gives the new physics permitting one to choose the correct
solution within the one parameter family.

\section{Deflagrations {\em vs.} detonations}
\label{sec:defldeto}

There are two kinds of bubbles allowed by the hydrodynamics.  These are
called deflagration and detonation bubbles~\cite{gyulassy} according to
the nature of the phase transition front.

Consider fluid flow in the rest frame of the phase transition front.
The incoming flow velocity is denoted by $v_1$ and the outgoing
velocity is denoted by $v_2$ (see \fig1). In a deflagration the
incoming flow is subsonic, $v_1<c_s$, and the fluid is accelerated by
the phase transition, $v_2>v_1$, whereas for a detonation the opposite
is true: $v_1>c_s$ and $v_2<v_1$.  Depending on whether the outflow is
sub- or supersonic, these processes are further divided into weak
($v_2<c_s$), Jouguet ($v_2=c_s$), and strong ($v_2>c_s$) deflagrations,
and strong ($v_2<c_s$), Jouguet ($v_2=c_s$), and weak ($v_2>c_s$)
detonations~\cite{courant}.

Consider then the structure of the bubble in the rest frame of the
ambient fluid.  When the bubble has grown large enough, any memory of the
initial shape of the nucleated bubble should be lost.  The bubble can
then be described as a similarity solution of the hydrodynamical
equations, i.e., it expands linearly with time, otherwise maintaining
its shape and profile. The fluid has to be at rest both at the center of the
bubble, and far away. Thus, in a deflagration bubble, the phase
transition front is preceded by a shock wave which heats up the fluid and
sets it moving outward. The phase transition front then brings the fluid back
to rest (see \fig2).
In a detonation bubble, the fluid is at rest when it is hit by the phase
transition front, which leaves the fluid flowing outward.  A rarefaction
wave follows, bringing the fluid at rest (see \fig3).
We denote the velocity of the phase transition front in this frame by
$\vdef$ for deflagrations and by $\vdet$ for detonations.
Weak deflagrations have $\vdef<c_s$, whereas for strong deflagrations
$\vdef>c_s$.  Detonations always have $\vdet>c_s$.  We can exclude
strong detonations~\cite{steinhardt}, because they leave the fluid
flowing too fast:  no such similarity flow
exists that would bring the fluid at rest at the center of the
bubble~\cite{hannu2}.
For a deflagration bubble, the velocity $\vs$ of the shock front is also of
interest.  In 1+1 dimensions the fluid flows at a constant velocity
$\vf$ between the shock and phase transition fronts.

For deflagrations, the velocities $\vd$, $\vf$ and $\vs$ are related
to the velocities $v_1$ and $v_2$ by the equations
$$
\begin{array}{lcl}
\vs =v_1|_{\rm shock}  &\mbox{\hspace{1cm}} &
\vd =v_2|_{\rm phase\; wall}
\\
\vf= \left.\frac{v_1-v_2}{1-v_1v_2}\right|_{\rm shock} &\mbox{\hspace{1cm}} &
\vf= \left.\frac{v_2-v_1}{1-v_2v_1}\right|_{\rm phase\; wall}  \,\, ,
\end{array}
$$
where the words ``shock'' and ``phase wall'' indicate the front at
which $v_1$ and $v_2$ are measured. For the phase transition front, we
sometimes use the notations $v_1=v_q$ and $v_2=v_h$, and
$T_q$ and $T_h$ for the temperatures of the incoming and outgoing
fluids, respectively.
The temperature between the fronts is then $\Tq$ and the temperature
of the hadron phase is $\Th$ (see \fig2).

The initial condition, matter at rest in the q phase, at temperature
$T_f < T_c$, and the equations of state of both phases, do not fix the
rate of bubble growth ($\vdef$ or $\vdet$), or the temperature
inside the bubble.  For simplicity, we illustrate this with the bag
equation of state\footnote{Usually the bag constant $B = L/4$ appears on
the q side, with the opposite sign.  This normalization of the
zero-point of energy does not affect the hydrodynamics.}
\beq
\begin{array}{lcl}
p_h(T) = a_h T^4 + L/4 & \mbox{ } & p_q(T) = a_q T^4\\
\epsilon_h(T) =  3a_h T^4 - L/4 & \mbox{ } & \epsilon_q(T) = 3a_qT^4
\end{array}
\label{bageos}
\eeq
which one gets from eq.~\ref{eos} by making the small-supercooling
approximation
\beq
   B(T) \sim \frac{L}{4} \left( 1-\frac{T^4}{T_c^4}\right).
\label{B(T)}
\eeq
Then $a_q = a$, $a_h = a - L/4T_c^4$.

The conservation of energy and momentum, and the non-negative entropy
production at the phase transition front restrict the possible values of
the incoming ($\epsilon_1$) and outgoing ($\epsilon_2$) energy densities
\cite{gyulassy} (see \fig4).
Detonations require a certain amount of supercooling.  If the latent
heat $L$ is large, the required supercooling can be quite substantial,
and the h matter then at a highly superheated state immediately behind the
phase transition front.  This has led to the conclusion that
deflagrations are the more likely process in the QCD phase transition in
the early universe~\cite{gyulassy}.

However, if the latent heat is small, detonations require less supercooling.
The nucleation temperature can be estimated~\cite{keijo} from
\beq
1-{T_f\over T_c}={A\over \sqrt{
171-4\ln(171^{3/2}/A)}},
\la{Tf}
\eeq
where 171 = $4\ln (t_cT_c)$ and
$A=\sqrt{16\pi/3}(\sigma^{3/2}/L \sqrt{T_c})$.
The values of the parameters $\sigma$ and $L$ are essentially unknown,
although
results from lattice calculations can be employed in giving rough estimates
for them. In \fig5
we show the region in the $(L, \sigma)$ parameter plane, where weak or
Jouguet detonations would be allowed.

To choose among the allowed solutions (see \fig4)
the internal mechanism of the
phase transition front needs to be considered.  This is the purpose of
our model presented in Section~\oldref{sec:model}, with the additional
parameter $\Gamma$, which will pick a single solution.  In the next
sections we turn to numerical results obtained for this model.

For the QCD phase transition we used the ``\qh'' parameters
$$
\begin{array}{lcl}
L=2 & \mbox{ } &\sigma = 0.1  \\
\xi_c = 1 & \mbox{ } & g_* = 51.25 \\
\end{array}
$$
Here and in the following all quantities are expressed scaled with
powers of $T_c$ to make them dimensionless: $L=L/T_c^4,\sigma=
\sigma/T_c^3, \xi_c=\xi_cT_c$. From
eq.~\ref{Tf}, these parameters correspond to the nucleation
temperature $\Tf = 0.9943$.
The values of the latent heat $L$ and the surface tension $\sigma$ are
suggested by pure glue lattice Monte Carlo
simulations~\cite{qcdpax,grossmannlaursen,iwasakikodisleo}.
The parameter $\xi_c$ shows up
neither when the nucleation temperature is calculated (in the
thin wall approximation, eq.~\ref{Tf}) nor when the steady-state variables
are calculated (in eqs.~(\oldref{2or}--\oldref{v(G)})).
However, it determines the
thickness of the phase transition surface. Because the
transition is only weakly first order, the actual $\xi_c$ might be
larger than our value. Notice that in addition to strongly interacting
degrees of freedom, the parameter $g_*$ includes weakly and
electromagnetically interacting degrees of freedom. Because the mean free
paths of these particles are much larger than those of strongly
interacting particles, these degrees of freedom are actually not active
during the early stages of the phase
transition~\cite{pantano,bonomettopantano}.
However, since we are mostly interested in the final stationary stages
of the phase transition, all the degrees of freedom are included.
For these parameters we expect deflagrations only (see
\fig5).

In the EW case, we assumed that $\alpha _W\approx 1/30$,
$m_{\rm top} = m_W$ and \mbox{$m_H\approx 40$}~GeV.
The small effective Higgs mass is necessary
in order to allow for a generation~of~the baryon asymmetry. Using the
improved effective potential~\cite{linde2}~we get our ``\ew'' parameters
$$
\begin{array}{lcl}
\alpha =0.0162 & \mbox{ } &\gamma = 0.1309 \\
\lambda = 0.0131 & \mbox{ } & g_* = 106.75 \\
\end{array}
$$
This corresponds to the nucleation
temperature $\Tf = 0.9957$~\cite{eikr}.

For the above {\qh} parameters
the lowest temperature $T_0$ where the symmetric minimum
$\phi =0$ still exists is $T_0=0.8771$. In the {\ew} case we have
$T_0=0.9828$. That these numbers are not too low indicates that our use
of the quartic effective potential $V(\phi ,T)$ is
justified \cite[eq.~(2.21)]{eikr}.

\section{Time-dependent phenomena}
\la{sec:time-dep}
To integrate \eqs\ref{yhtalot} we wrote a simple 1+1 dimensional relativistic
hydrodynamics code following Wilson \cite{hannu1,wilson}.
Thus we use explicit
differencing with operator splitting for the hydrodynamic equations.
The code variables are $\phi$, $\pi \equiv \partial_t\phi$, $E \equiv
\gamma\bigl[3aT^4 + V(\phi,T) - T{\partial V\over\partial T}]$, and
$Z \equiv \gamma^2v\bigl[4aT^4 - T{\partial V\over\partial T}]$.
The velocity $v$ is solved from $E$ and $Z$.  The temperature $T$ (for
each grid point) is solved from $E$ and $\phi$ using the functional form
of $V(\phi,T)$.  This value for $T$ is then used in $V(\phi,T)$ for
evolving $\phi$.  The transport terms for $E$ and $Z$ are handled with
the FCT method \cite{boris,kataja}.

We use reflective boundary conditions (see \fig6). The center of the
initial ``bubble'' of new phase is placed at one end of the grid.  Allowing
the moving wall to reach the other end simulates the collision with
another similar bubble.  The code corresponds to a planar geometry, so
these are not true spherical bubbles.
In this paper we study
the motion of a planar phase wall.

\subsection{Initial conditions}
Before starting the actual integration,
the initial conditions have to be specified. The initial bubble
has to be larger than critical to start growing, but it is not clear to what
extent there is a fluctuation in the
temperature associated with the fluctuation
in the order parameter. Possibly the temperature is a bit higher near the
critical bubble than farther away from it, because latent heat is released
in the formation of the critical bubble.
We can estimate typical temperature fluctuations with classical
fluctuation theory \cite{lanstat}: for the quark matter equation of state
\beq
\frac{\langle (\Delta T)^2 \rangle}{T^2}=\frac{1}{12aT^3V}\,\, .
\eeq
For the {\qh} case, with a radius $R_c\approx 5\xi_c$ of the critical
bubble, we get $\Delta T \approx 0.005$ in units of $T_c$. Comparing
with $T_f$ in the previous section, this is very large. For the {\ew}
case we get $\Delta T \approx 0.001$ \cite{krs}. However, we think that
it is most straightforward and in the spirit of the nucleation
calculation (\eg in ref.~\cite{eikr}) to assume that the temperature inside
the initial bubble is just $\Tf$. What is most important, the details
of the initial bubble have no effect on the final
steady-state configuration and the asymptotic variables
$\Tq$, $\Th$, $\vs$, $\vf$ and $\vd$ (or $\vdet$),
if only the nucleated bubble starts growing.

Assuming as initial conditions that the fluid velocity
vanishes everywhere and that the temperature is constant and equal to $\Tf$,
we still have to decide the shape and size of the initial bubble.
These variables have some significance during the early stages of
bubble evolution, since they affect the initial shape
of the shock front and also its initial acceleration.
In $1+1$ dimensions it is possible to analytically find the
extremum bubble of the effective action by solving
the equation $\phi ''(x)=\partial _{\phi}V(\phi , \Tf)$.
With $M^2\equiv\gamma (\Tf ^2-T_0^2)$, $\delta\equiv\alpha\Tf$ and
$\bar{\lambda}\equiv 9\lambda M^2/2\delta ^2$
the solution is
\beq
  \phi (x) = \frac{2\delta}{3\lambda} \,
             \frac{ 1 - \sqrt{1-f(x)} }{ f(x) } \, \bar{\lambda},
\eeq
where $f(x)\equiv (1-\bar{\lambda}\coth ^2Mx)/(1-\coth ^2Mx)$.
A  bubble obtained from this by increasing both the amplitude and
radius by a small factor (by 5\%) was normally used as the initial
configuration. We experimented also with the exact extremum bubble and
with a bubble smaller than this one. The exact extremum bubble did not
evolve anywhere and the subcritical bubble collapsed, leaving behind
a disturbance in the temperature and flow velocity propagating
outward with sound velocity. This disturbance is caused by the fact
that matter has to flow inward to fill in the area of lower energy density
from where the subcritical bubble has disappeared.

\subsection{Numerical results}
In \fig7 the initial stages of bubble formation are shown
for the {\qh} parameters. Immediately after the nucleation, a shock front
is originated which spreads out information about the nucleation.
At first the shock front is not sharp.  The temperature
starts to rise inside the bubble and very soon the phase transition
front begins to get shape. At about the moment $t=100 (1/T_c)$,
both the phase transition front
and the shock front are clearly visible.

To get a more precise picture of the phase transition surface,
we can use eqs.~(\oldref{stat1}--\oldref{stat3}).
(The solution of these eqs.\ is discussed in Section~\ref{sec:stationary}).
In \fig8 the order parameter
$\phi (x)$, the temperature $T(x)$, the velocity $v(x)$ and
the quantity $\doo _T V$ are shown in the rest frame of
the phase transition surface. Hadron phase is on the left and
quark phase on the right. The width of the surface layer
is a few correlation lengths. The curves resemble the
$\tanh (x)$-function but for some other parameters the resemblance
is not as clear. Specifically, the temperature and velocity
distributions lose their symmetry and are shallower
on the hadron side where the effective potential $V(\phi ,T)$
is non-zero. The ``center points'' of these distributions
are not quite at the same place as that of the order parameter,
but are shifted towards the hadron phase. For some parameters
the shift can be of the order of $\xi_c$.

In \fig9 the development of the shock front is illustrated.
The shock front is shown at times $t=160$ and $t=3840$. At early times,
the shape of the initial configuration strongly affects the shape
of the shock front. However, after some time the shock front
sharpens to a discontinuity \cite{hannu2} irrespective of its initial shape.
To understand the physical reason for this, consider yourself moving
with the shock front and looking back towards the heated quark matter.
Particles farther away from the front recede more slowly than particles
just at the front. When the shock front is still smooth, it follows
from energy-momentum conservation that entropy is conserved, i.e., with
the quark matter equation of state,
\beq
\doo _t(T^3\gamma )+\doo _x(T^3\gamma v)=0 \,\, .
\eeq
This means that the temperature has to rise farther away from the front
in order to accommodate the entropy of the matter moving with a lower
velocity. However, as another consequence of energy-momentum conservation,
temperature cannot rise enough to accommodate all the entropy inside
a constant-sized volume, and
a ``traffic jam'' -phenomenon occurs causing a discontinuity.
An upper limit to the rate of jamming is clearly given by the
difference of the velocity of the matter going into the jam
and the velocity of the jam. This difference is just $\vf$.
Then the time scale of the sharpening is determined by
$\vf$ and on how smooth the shock front was in the beginning.
The latter depends on how near the initial configuration was of
the extremum bubble. Because the shock front can initially be
very wide and the flow velocity is very small,
the time scale of the sharpening is very large.

In \fig10 a collision of two bubbles is shown. Due to the
use of reflective boundary conditions, our grid corresponds to
a situation in which several bubbles nucleate simultaneously at
equal spacings, and therefore collisions are possible. The distance
between the bubbles is in our picture $\Delta x=640$ which is much
less than the actual distances in the early universe, but this
is not essential for the present analysis. At time $t=240$,
both the phase transition surface and the shock front are
moving to the right. At $t=720$, the shock fronts of
neighbouring bubbles have collided, and the quark matter
between the reflected shocks heats to a temperature higher
than $\Tc$. At $t=960$ the shock front and the phase transition
surface are just about to collide; at $t=1040$, the collision
has just happened. The temperature of the hadron phase increases
significantly. A shock front~(a) continues to the hadron phase
heating it up and making the matter move, but at the same time
a rarefaction wave~(b) is reflected back to the quark phase, cooling
it down. The phase transition surface continues to move to the
right, but its velocity has decreased from $0.06$ to about
$0.015$. Some simple scenarios for the collision of a shock front
and a deflagration front have been presented in ref.~\cite{kk},
and the course of events in \fig10 is a combination
of the scenarios~{\bf a} and~{\bf c}.

As was seen in \fig10, even one shock front
can heat the hadronic matter considerably. This implies that
only a few shock fronts are needed to raise the temperature of
the hadronic matter to $\Tc$. This situation
is shown in \fig11,
where several collisions are allowed to happen.
The upper picture shows $T(t,x)$ and the lower
picture $\phi (t,x)$. As the temperature
of the hadron matter rises to $\Tc$ the growth of the hadron bubbles
is halted. In the cosmological context, a stage of slow growth
of the hadron bubbles along with the expansion of the universe
follows. In the electroweak case, the latent heat is smaller
and the critical temperature is not reached. The bubbles
fill the space and the fluid reaches the reheating temperature $\Tr$.
However, we must remember that the present simulations are 1+1~dimensional.
In the 3-d case, the shock fronts are considerably
weaker~\cite{hannu2} and the time it takes for the
universe to reheat is not necessarily the same as in the 1-d case.
Neglecting the expansion
of the universe, the fraction of space
taken by the hadron matter immediately after reheating in the QCD case
and the reheating temperature in the EW case are nevertheless
the same as in the 1-d case.

All the numerical solutions we have discussed so far were deflagrations.
However, using a larger supercooling than eq.~\ref{Tf}
indicates for the above ({\qh}) values of $\sigma$ and $L$,
we found detonation solutions.  For $\Tf =0.90$ such a solution
is shown in \fig12.
The upper picture shows $T(t,x)$ and the lower
picture $\phi (t,x)$. The detonation expands with velocity $v_{\rm
det}=0.90$ and a rarefaction front follows slightly behind. This
is a weak detonation.

As was seen in \fig5,
for some values of $\sigma$ and $L$ even the physical nucleation temperature
is so low that detonations could appear.
In particular, even though the latent heat $L$
is usually taken to be of the order of $T_c^4$, there is nothing
in the lattice computations to rule out the possibility that $L$ could
as well be $0.1T_c^4$ or even less. Using the parameter values\footnote
{The larger value of $\xi_c$ is needed for the potential of eq.~\ref{V}
to be applicable for the range of temperatures in question.}
$\sigma = 0.1$, $L = 0.1$, and $\xi_c = 6$, we did a sequence of
computer runs varying the value of $\Gamma$ (see \fig13).
This results in considerable supercooling, and the nucleation temperature
(eq.~\ref{Tf}) is $T_f = 0.891T_c$.
For small values of $\Gamma$ the bubbles grow as weak deflagrations.
Increasing $\Gamma$ increases $\vdef$, until at about $\Gamma = 10$
it approaches $c_s$. Now $\vdef > c_s$ would indicate a strong
deflagration. Instead, as $\Gamma$ is further increased,
the solution shifts to a weak detonation.

\section{Steady-state variables of deflagration bubbles}
\label{sec:stationary}

Very soon after the nucleation, the growing bubbles reach a
steady-state configuration~(see \fig7).
In the steady-state situation
there exists a simple and very accurate way of finding out the
interesting stationary variables $\Tq$, $\Th$, $\vs$, $\vf$ and $\vd$
of the deflagration bubble apart from the above-presented
integration of \eqs\ref{yhtalot}. This method
can also be used to check the accuracy of the above integration.
In the rest frame of the phase transition front the equations
to be solved were given in eqs.~(\oldref{stat1}--\oldref{stat3}).

Let us think that with the two conservation
equations \ref{stat2} and~\ref{stat3}, we solve
for $T(x)$ and $v(x)$ in terms of $\phi (x)$ and $\phi '(x)$.
Substituting
these to eq.~\ref{stat1}, we get a second order differential
equation for the field $\phi$ alone. However there are {\em three}
boundary conditions: the derivative $\phi '(x)$ must vanish
asymptotically
in both phases and we know the value of the field $\phi (x)$ (only) in
the quark phase. The system is over\-determined, and only for certain
values of the ``constants'' are there solutions.  This is thus an
eigenvalue problem.
Assuming that
we are given $\Tq$, we can solve for $\Th$, $\vd$ and $\vf$.

Next, consider the shock front. Because the shock front is
in the quark phase, the field $\phi$ vanishes everywhere and we are left
with very simple energy-momentum conservation equations. Solving
them \cite[\S 135]{landau} we get
\beq
\left.
\begin{array}{l}
v_1 = \frac{1}{\sqrt{3}}\sqrt{\frac{3\Tq ^4+\Tf ^4}
{3\Tf ^4+\Tq ^4}}\\
v_2 = \frac{1}{3v_1} .
\end{array}
\right. \label{sh}
\eeq
Therefore, given $\Tf$ and $\Tq$, we can write down $\vs =v_1$
and $\vf =(3/2)(v_1-v_2)$.

Now remember that {\em a priori} we only know $\Tf$. However,
guessing some $\Tq$, we get both from eqs.~(\oldref{stat1}--\oldref{stat3})
and eq.~\ref{sh} a
value for $\vf$. When we manage to guess such a $\Tq$ that these
two numbers agree the whole problem is solved.
It is easy to make this method of solving the
steady-state variables very accurate. Therefore we can use this
method to check the accuracy of the dynamical integration
with all time derivatives. With a reasonable number of grid points,
the differences between the results of the two methods of
integration on the steady-state variables
are much less than 1\% . Using this method, we can also
easily calculate the steady-state variables as a function of~$\Gamma$.

\subsection{Analytical approximations}
\label{sec:anal.appx}
In this section we study what can be said analytically of the solutions
of the steady-state equations~(\oldref{stat1}--\oldref{stat3}). We use
the bag equation of state (\oldref{bageos},\oldref{B(T)})
and assume that the
velocities $\vq$ and $\vh$ are non-relativistic ($v^2\ll1$)
and that the temperatures
$\Tq$ and $\Th$ are near $\Tc$ ($\Tq-\Th\ll\Tc$).
The entropies are $s_q=4\aq\Tq ^3$ and $s_h=4\ah\Th ^3$
and the enthalpy is $w=Ts$.
The one-parameter family of solutions of the two equations \ref{stat2}
and \ref{stat3} has been studied in detail in the literature
\cite{gyulassy,hannu2,kk,eikr,keijo,hkllm},
and the main
problem is to find which solution the new equation \ref{stat1}
picks out of this family.

Evaluating eqs.~\ref{stat2} and~\ref{stat3} in the rest frame of
the phase transition front for $x=-\infty$ and $x=\infty$
one obtains the usual
energy-momentum conservation equations
\beq
\begin{array}{l}
\wq\vq =\wh\vh \\
\wq\vq ^2+\pq =\wh\vh ^2 + \ph  \,\, .  \la{sai}
\end{array}
\eeq
\mbox{}From the upper equation it follows that
\beq
\frac{\vh}{\vq}=\frac{\wq}{\wh}=\frac{\aq\Tq ^3}{\ah\Th ^3}\approx
\frac{\aq}{\ah}\equiv r
\,\, . \la{appro1}
\eeq
Thus the fluid velocity is related to the deflagration front velocity
by
\beq
\vf =\vh -\vq =(1-1/r)\vd
\,\, . \la{vfluid}
\eeq
\mbox{}From the lower of \eqs\ref{sai} we get
\beq
\vq\vh =r\vq ^2=\frac{\ph -\pq}{\wh -\wq}=\frac{\ah\Th ^4+L/4-\aq\Tq ^4}
{4\ah\Th ^4 -4\aq\Tq ^4} \la{appro2}
\eeq
which in the limit $\Tq ,\Th\approx T_c \equiv 1$ implies that
\beq
1-\Th =r(1-\Tq )+r(r-1)\vq ^2
\,\, . \la{appro3}
\eeq

To find the consequences of eq.~\ref{stat1}, multiply
it by $\phi '(x)$ and integrate over
the real axis. Using the equations
$\partial{V(\phi ,T)}/\partial{\phi}={dV}/{d\phi}-
(\partial{V}/\partial{T})({dT}/{d\phi})$, $V(-\infty)=-L(1-\Th)$,
$V(\infty)=0$ and $\phi'(\pm\infty)=0$, and replacing the velocity $v(x)$
by its absolute value (see \fig8) one obtains
\beq
L(1-\Th )= {1\over\Gamma}\int_{-\infty}^\infty [\phi'(x)]^2v(x)\,dx
-\int_{\Tq}^{\Th}\oder{V}{T}dT
\,\, .\la{velo}
\eeq
This is easy to solve in the limit $r=\vh/\vq\to1$ and
$\Th,\Tq\to\Tc$ since then one can take $v(x)\approx\vh\approx\vq$
out of the first integral
(which then gives the interface tension~$\sigma$) and neglect
the last term. The result is
\beq
\vh\approx\vq\approx v_b\equiv\Gamma{L\over\sigma}(1-\Tq).
\label{vb}
\eeq
This is the formula for bubble wall velocity derived earlier
\cite{turok,linde2,keijo,liu} for the case of small change
in flow velocity, which may be appropriate for the EW phase transition.

In the more general case $r>1$ note first that, because $\phi'(x)$
is approximately symmetric around $x=0$, the first term in eq.~\ref{velo}
can be approximated by $(\sigma/\Gamma)\cdot (\vq+\vh)/2$.
\mbox{}From \fig8 one sees that at $x=0$,
$\doo V/\doo T$ is about $L/2$ (in the hadron phase, $\doo V/\doo T
\approx L$, since $V(-\infty )\approx -L(1-\Th )$, see above).
We therefore approximate the second term
by $-\kappa (L/2)(\Th -\Tq)$,
where $\kappa$ is of order unity. Using \eqs\ref{appro1}
and \ref{appro3}, eq.~\ref{velo} then becomes
\beq
(1-\frac{\kappa}{2})Lr(r-1)\vq ^2-\frac{\sigma (r+1)}{2\Gamma}\vq
+L(1-\Tq)[r+\frac{\kappa}{2}(1-r)]=0
\,\, .\la{2or}
\eeq
\mbox{}From this one can solve $\vd =\vh =r\vq$ as a function of $\Tq$.
If further $\kappa =1$, eq.~\ref{2or}
simplifies into the form
\beq
\left(\frac{r-1}{r+1}\right)\vd ^2-\frac{\sigma}{L\Gamma}\vd+
r(1-\Tq )=0
\,\, ,\la{vdef}
\eeq
which for small $\Gamma$ (small $\vd$) again gives
the result $\vq =v_b$ in eq.~\ref{vb}.

It is worth noticing that from eq.~\ref{2or} we get
a lower bound for $\Tq$. Namely, we know that eq.~\ref{2or} has
a solution and this gives
\beq
\Tq \ge 1-\frac{\sigma ^2(r+1)^2}{16L^2\Gamma ^2r(r-1)(1-\kappa /2)
[r+(\kappa /2)(1-r)]}
\,\, .
\eeq
The numerical value of this formula is useful for large $\Gamma$.
However, even for
moderate $\Gamma$ it is essential
to notice the {\em existence} of a lower bound: the temperature
$\Tq$ is not a free physical parameter, and when the whole
expanding physical bubble is considered, the shock wave always
heats quark matter just enough to reach the safe $\Tq$ area.

To obtain the result for a true deflagration bubble one finally has
to eliminate the temperature $\Tq$ from eq.~\ref{vdef} by using
eq.~\ref{sh}. Expanding $\vf$ in powers of $1-\Tq$ and $1-\Tf$
one gets
\beq
\vf = \sqrt{3} \, [(1-\Tf)-(1-\Tq)] \,\, .
\eeq
Using eq.~\ref{vfluid} and substituting $1-\Tq$ in terms of $\vd$
into eq.~\ref{vdef} gives
\beq
\left(\frac{r-1}{r+1}\right)\vd ^2-\left(
\frac{\sigma}{L\Gamma}+\frac{r-1}{\sqrt{3}}\right)\vd +
r(1-\Tf )=0 \,\, .
\eeq
\mbox{}From this the final equation for $\vd$ as a function of
$\Gamma$ is
\beq
\vd=\left[\frac{\sigma}{L\Gamma}+\frac{r-1}{\sqrt{3}}
-\sqrt{\left(\frac{\sigma}{L\Gamma}+\frac{r-1}{\sqrt{3}}\right)^2-
4r(1-\Tf)\left(\frac{r-1}{r+1}\right)}\right]
\left/2\left(\frac{r-1}{r+1}\right) \right.
\,\, .\la{v(G)}
\eeq
This is an excellent approximation for non-relativistic
deflagration front velocities as will be seen in the next section
(see \fig15).
In the limit $\Gamma\to 0$ eq.~\ref{v(G)} correctly reduces to the
further approximation $\vd \approx r\Gamma (L/\sigma )(1-\Tf)$.

\subsection{Numerical results for steady-state walls}
\subsubsection{The quark-hadron phase transition}
In \fig14 the temperatures $\Tf$, $\Tq$ and $\Th$
are shown as a function of $\Gamma$ for {\qh}
parameters. Small $\Gamma$ means
large friction and small velocities; large $\Gamma$ means
small friction and large velocities. When the deflagration front velocity
is small, equation~\ref{vfluid} tells us that the fluid velocity
is very small (for our present parameters, $r\approx 1.1$ so that
$\vf\approx 0.1\vd$). Then from \eqs\ref{sh} we learn that $\Tq$
has to be very close to $\Tf$. For large $\Gamma$ the situation is opposite:
the velocities are larger and $\Tq$ is higher.
The temperatures $\Tq$ and $\Th$
satisfy the relation \ref{appro3} very accurately. Notice that entropy
production at the shock front requires the condition
$\Tq >\Tf$ and obviously one also has to obey the condition $\Th <\Tc$,
but nothing prevents $\Tq$ from exceeding $\Tc$. This fact has been
noticed before (see \eg ref.~\cite{gyulassy}), but usually it is not taken
seriously. One reason may be that some authors
neglect proper boundary conditions  and thus confuse
$\Tq$ with $\Tf$.
Another is that it has been argued in
refs.~\cite{ruuskanen,gorenstein} that a transition front between
quark matter at the temperature $\Tq >\Tc$ and hadron matter
at the temperature $\Th <\Tc$ is impossible, basically because
such a front would be mechanically unstable. However, both arguments
are based on the assumption that at $T_c$ there
exists a homogeneous mixed phase, and that in the
transition zone between the quark phase
and the hadron phase, the state of the matter is at some point
just in this homogeneous mixed phase.
Then the transition front would be equal to
two transition fronts, one from the quark phase to the mixed phase and
the other from the mixed phase to the hadron phase.
If there is a microscopic order parameter
field --- like in our model ---
the order parameter interpolates between the two minima of the
effective potential in the transition zone,
and there is no homogeneous mixed phase.
The only mechanism by which a phase transition surface of this
kind could in principle split is that a rarefaction wave detaches from it,
and the analysis in ref.~\cite{gorenstein} does not
apply to rarefaction waves. Hence, there seems to be no reason why
$\Tq$ could not exceed $\Tc$, if the phase transition
effectively includes an order parameter field.

In \fig15 the propagation velocity of the phase
transition surface $\vd$ is shown with solid line. With the dashed
line we have drawn the deflagration front velocity from eq.~\ref{v(G)}.
This equation is seen to hold very well when the velocity
$\vd$ is non-relativistic. The simple small-velocity
approximation $\vd \approx r\Gamma (L/\sigma )(1-\Tf)$ is drawn
with dotted line. In \fig16
the fluid velocity $\vf$
is shown as a function of $\Gamma$ (solid line)
and in \fig17 the shock velocity $\vs$ is drawn (solid line).
The shock velocity is compared to the sound velocity.

In \fig16, the dotted line shows the entropy production.
By entropy production
we mean the relative change of the total entropy of a fluid element
as the shock front and the phase transition surface sweep over it.
One must note that the volume of the fluid element changes in the
course of the process by the relative amount (see \fig18)
\beq
\rho _V=\frac{V_{\rm final}}{V_{\rm initial}}=
\frac{1-\vf /\vs}{1-\vf /\vd}>1 \,\, .
\eeq
Therefore we define the entropy production by
\beq
\Delta s\equiv \frac{\sh (\Th )\rho _V-\sq (\Tf ) }{\sq (\Tf )} \,\, .
\la{ent}
\eeq
In the quark-hadron phase transition
where there is a conserved baryon number,
this is just the relative change of entropy per baryon.
The entropy production in eq.~\ref{ent}
is related to the quantity $\Delta s_{\rm pw}\equiv
\sh \gamma _h\vh -\sq \gamma _q\vq$ (measured in the rest
frame of the phase transition surface),
which is often (\eg in ref.~\cite{eikr})
used to describe entropy production, and to the analogously defined
quantity $\Delta s_{\rm sh}$ measured in the rest frame of the shock front,
by the equation
\beq
\Delta s=\left(\frac{\Delta s_{\rm sh}}{\gamma_{\rm sh}v_{\rm sh}}
+\frac{\Delta s_{\rm pw}}{\gamma _{\rm def}\vd}\rho _V\right)/\sq (\Tf)
\,\, .
\eeq
The first term in the numerator is vanishingly small in comparison
to the second.

According to ref.~\cite{hkllm} the quantity $\eta \equiv -\Tc
(d\vd /d\Tq )\vd$ determines the stability
of expanding bubbles. If $\eta > 1$, the bubbles are stable
at all length scales; if $\eta < 1$, large scale fluctuations
are unstable. For our {\qh} parameters, the quantity $\eta$ is
drawn in \fig17 with dotted line.
While the analysis of ref.~\cite{hkllm} is not
suited for large $\Gamma$ where $\Tq$ can exceed $\Tc$,
we notice, however, that for $\Gamma < 0.83$ our numerical
results imply $\eta$ to be less than unity and therefore
large scale fluctuations should be unstable in that case.
It would be interesting to expand the present code to include more space
dimensions to see whether the expanding bubbles remain stable.

\subsubsection{The electroweak phase transition}
In figs.~19--20 we address the same questions as above but for the
{\ew} parameters. Qualitatively the behaviour is rather similar to what
it was for the {\qh} case. However, some differences worth a comment
exist. The most important difference is that the electroweak phase transition
is essentially a massless transition: the number of effective degrees of
freedom changes very little at $T_c$. Quantitatively this is expressed
by the fact that $r=a/(a-L/4)=1.0018$ is much closer to unity than in
the {\qh} case. Because of the smallness of $r$, there is a new temperature
relevant for the phase transition. This is the reheating temperature $\Tr$
(in the abrupt reheating scenario), defined by
the equation $\epsilon _q(\Tf )=\epsilon _h(\Tr )$. For our
present {\ew} parameters, $\Tr =0.9965$.

Consider the temperatures (\fig19) and the deflagration
front velocity (\fig20).
Because the energy liberated in the phase transition is travelling in
the compressed region behind the shock front, the temperature $\Th$ can
never reach $\Tr$. This keeps $\Th$ low even for large $\Gamma$.
On the other hand, it follows from eq.~\ref{v(G)} that
for large $\Gamma$ the velocities $\vd$ and $\vf$
grow large (now $\vf\approx 0.002\vd$). Then it is seen
from \eqs\ref{sh} that the temperature $\Tq$ has to rise
considerably when $\Gamma$ is large. But with $r$ close to unity
and the difference between $\Tq$ and $\Th$ large, it is seen from
eq.~\ref{appro3} that the deflagration front
velocity has to be very large indeed for large $\Gamma$.
In fact the velocity $\vd$ becomes moderately
relativistic and eqs.~(\oldref{sai}--\oldref{v(G)}) are
no longer strictly applicable.
Another way to understand the connection between the high temperature
$\Tq$ and the large velocity $\vd$ is to notice that the closer
$\vd$ is to the shock front velocity, the thinner is the area
between the shock front and the phase transition surface.
Therefore this thin area has a high temperature in order to
accommodate all the latent heat released.

Comparing figs.~14 and~19
we notice that for the {\ew}
parameters the values of $\Gamma$
where $\vd$ changes rapidly are larger than
for the {\qh} case. From eq.~\ref{v(G)} we see that
the relevant scale for $\Gamma$ is roughly $\sigma (r+1)/L(r-1)$.
For the {\qh} case the numerical value of this quantity is
$1.1$ and for the {\ew} case $96$ which explains the difference.

Finally, let us compare our results to those of some other authors.
Recently, the velocity of growing deflagration bubbles in the
EW phase transition has been estimated
for instance in refs.~\cite{linde2,liu}.
In ref.~\cite{liu} the authors note that the relation $\vd > 1/30$
has to be satisfied in order not to diffuse away
the baryon asymmetry and that probably $\gamma v\approx 1-2$,
that is, $v\approx 0.7-0.9$. This
corresponds to strong deflagrations, which seem to be an unlikely
mechanism for bubble growth~\cite{landau,courant}.
In ref.~\cite{linde2} velocities of order $0.1-0.2$ are obtained.
Using simple kinetic theory to estimate the value of~$\Gamma$ we get
$\Gamma\approx 10-100$~\cite{turok,ck}, and then
from \fig20 mildly relativistic velocities
appear to be very natural.

\section{Conclusions}
\label{sec:concl}

We have presented a model for phase transition bubbles in the early
universe. In our model an order parameter field $\phi$ with an effective
potential $V(\phi,T)$ is coupled to a fluid with a dissipative constant
$\Gamma$.  Starting from an initial condition of a newly nucleated
bubble we have numerically evolved the coupled hydrodynamical and field
equations to follow the growth of the bubble in 1+1 dimensions.  After
some time the bubble reaches a stationary (similarity) state, where it
grows at a constant velocity.
We have then also studied the solutions to the corresponding stationary
equations, both numerically and in analytical approximations.

Typically the bubbles grow as weak deflagrations, and therefore with a
subsonic velocity. The growth velocity
is determined by the value of $\Gamma$, a large $\Gamma$ (a weak coupling
between $\phi$ and the fluid) leading to a large velocity and vice versa.
Thus, we have been able to reduce the calculation of the growth
velocity to the microscopic calculation of $\Gamma$. Our results show that
the preheating caused by the shock front plays an essential role in
the growth process, and that the temperature~$\Tq$
could even exceed~$\Tc$. Reheating caused by
collisions of expanding bubbles was also explicitly computed.

If one uses simple kinetic theory to estimate $\Gamma$ for the EW
transition, one is naturally led to mildly relativistic velocities.
For the QCD transition, one would
dimensionally expect that $\Gamma\approx 1$ \cite{keijo}.
Then from \fig15 for $L = 2T_c^4$, $\sigma = 0.1T_c^3$
we  note that
$\vd\approx 0.06$. The expectation thus is that the velocities
in the QCD case are smaller than in the EW case.

In some regions of the parameter space the solutions switch
from weak deflagrations to weak {\em detonations}, as $\Gamma$ is increased.
It has been speculated that instabilities could turn expanding
deflagration bubbles into detonations \cite{kamion}.
We have  now found that in
some cases the bubbles could expand as detonations even from the beginning.
Often it has been assumed that detonations in these phase
transitions would have to be Jouguet detonations \cite{steinhardt,gyulassy},
as in chemical burning \cite{landau,courant}, but this appears not
to be the case \cite{laine}.  In the cosmological context, weak
detonations require less extreme conditions than Jouguet detonations,
making detonations more likely.

For the QCD phase transition, there is an interesting parameter region
that cannot presently be ruled out. Here the latent heat is rather
small, say $L = 0.1T_c^4$, and the surface tension is similar,
say $\sigma = 0.1T_c^3$.  Then the average distance between the
nucleation centers
would be of the order of 10 meters and large scale hadronic
inhomogeneities would result.  With our model we find that such bubbles
could grow as detonations, leading to a picture of the phase transition
that is rather different from the usual one.

\section*{Acknowledgements}
We thank the Academy of Finland for financial support,
K. Rummukainen and P.~V.~Ruuskanen for discussions and the Finnish
Center For Scientific Computing for computational facilities.

\newpage
\section*{Figure captions}
\setlength{\parindent}{0.0cm}
{\bf Fig. 1}:{  The velocities in the rest frames
of the phase transition and shock fronts.}
\tyhja

{\bf Fig. 2}:{  Structure of a deflagration bubble.}
\tyhja

{\bf Fig. 3}:{  Structure of a detonation bubble.}
\tyhja

{\bf Fig. 4}:{  The values of incoming (q) and outgoing
(h) energy densities at the phase transition front, corresponding
to deflagrations (upper left triangular region) and detonations (lower
right).  This figure is for a bag model with $r \equiv a_q/a_h = 2$.
Point A corresponds to $T_q = T_h = T_c$.
In deflagrations the fluid has been heated by the
shock, so $T_q > T_f$,  whereas for detonations $T_q = T_f$.
For a given initial temperature $T_f$, there is a one-dimensional space of
solutions, indicated by the dashed lines.
The requirement of non-negative entropy production restricts the solution
below the line $\Delta S = 0$.
Thus, in this case,  only deflagrations are allowed for $T_f = 0.9 T_c$.
For $T_f = 0.8 T_c$, both deflagrations and weak detonations are
possible.  For $T_f = 0.7 T_c$, Jouguet detonations are allowed, too.}
\tyhja

{\bf Fig. 5}:{  For sufficiently small latent heat $L$,
or sufficiently large surface tension~$\sigma$, detonations are
possible in addition to deflagrations.
The precise regions in the $(\sigma,L)$ plane depend on
the equation of state.  This figure is for a bag model with $a_q =
51.25\pi^2/90$, $a_h = a_q - L/4T_c^4$ (solid boundaries), or with $a_q
= a_h + L/4T_c^4$, $a_h = 17.25\pi^2/90$ (dashed boundaries).}
\tyhja

{\bf Fig. 6}:{  The meaning of the reflective boundary conditions.}
\tyhja

{\bf Fig. 7}:{  The early stages of bubble growth.
The upper picture shows $T(t,x)$ and the lower
picture shows $v(t,x)$ for the {\bf qh} parameters.}
\tyhja

{\bf Fig. 8}:{
The variables $\phi (x)$, $T(x)$, $v(x)$ and $\partial V/\partial T$
as measured in the rest frame of the stationary phase transition
surface. Hadron phase is on the left and quark
phase on the right. By the velocity $v$ we denote here the absolute value of
the velocity: the direction of flow is from the quark to the hadron phase.}
\tyhja

{\bf Fig. 9}:{
The sharpening of the shock front.}
\tyhja

{\bf Fig. 10}:{  The collision of two bubbles.
See the body of text for explanation.}
\tyhja

{\bf Fig. 11}:{
Final stages of the phase transition for the {\bf qh} parameters.}
\tyhja

{\bf Fig. 12}:{
A detonation solution for the {\bf qh} parameters with
$\Gamma =2$ and $T_f=0.90$.
Note that $\phi_{\rm min}$ depends on temperature.}
\tyhja

{\bf Fig. 13}:{  The bubble growth velocity for $\sigma =0.1$, $L=0.1$
and $\xi _c=6$ as a function of the parameter $\Gamma$. At
approximately $\Gamma =10$ the solution changes from a weak
deflagration to a weak detonation. }
\tyhja

{\bf Fig. 14}:{
The temperatures $T_f$, $T_c$, $T_q$, $T_h$
as a function of $\Gamma$ for the
{\bf qh} parameters. }
\tyhja

{\bf Fig. 15}:{
The velocity $v_{\rm def}$ as a function of $\Gamma$ for the
{\bf qh} parameters. The dashed line is the approximation from
eq.~\ref{v(G)} and the dotted line is the further approximation
$\vd\approx r\Gamma (L/\sigma )(1-\Tf)$.}
\tyhja

{\bf Fig. 16}:{
The fluid velocity and the entropy production
for the {\bf qh} parameters.}
\tyhja

{\bf Fig. 17}:{
The shock velocity and the parameter $\eta$
for the {\bf qh} parameters.}
\tyhja

{\bf Fig. 18}:{
The dashed lines indicate the average fluid flow in a similarity
deflagration solution. The volume of a fluid element
changes when the shock front and the deflagration front sweep
over it.}
\tyhja

{\bf Fig. 19}:{
The temperatures $T_f$, $T_c$, $T_q$, $T_h$ and
$T_{\rm reheat}$ for the {\bf ew} parameters.}
\tyhja

{\bf Fig. 20}:{
The deflagration front velocity, the fluid velocity and the
entropy production
for the {\bf ew} parameters.}

\end{document}